\documentclass[10pt,english,a4paper,showpacs,amsmath,amssymb,floatfix,twocolumn]{revtex4-1}
\usepackage{amsmath,amstext,amssymb,graphicx,babel,geometry,setspace,bm,dcolumn,color}
\begin{document}
 
\title{Transverse Ising Model: Markovian evolution of classical and quantum correlations under decoherence}
\author{Amit Kumar Pal}
\email{ak.pal@bosemain.boseinst.ac.in}
\author{Indrani Bose}
\email{indrani@bosemain.boseinst.ac.in}
\affiliation{Department of Physics, Bose Institute, 93/1 Acharya Prafulla Chandra Road, Kolkata - 700009, India}

\begin{abstract}
The transverse Ising Model (TIM) in one dimension is the simplest model which exhibits a quantum 
phase transition (QPT). Quantities related to quantum information theoretic measures like 
entanglement, quantum discord (QD) and fidelity are known to provide signatures of QPTs. The issue 
is less well explored when the quantum system is subjected to decoherence due to its interaction, 
represented by a quantum channel, with an environment. In this paper we study the dynamics of the 
mutual information $I\left(\rho_{AB}\right)$, the classical correlations $C\left(\rho_{AB}\right)$
and the quantum correlations $Q\left(\rho_{AB}\right)$, as measured by the QD, in a two-qubit state 
the density matrix of which is the reduced density matrix obtained from the ground state of the TIM in 
1d. The time evolution brought about by system-environment interactions is assumed to be 
Markovian in nature and the quantum channels considered are amplitude damping, bit-flip, phase-flip
and bit-phase-flip. Each quantum channel is shown to be distinguished by a specific type of 
dynamics. In the case of the phase-flip channel, there is a finite time interval in which the quantum 
correlations are larger in magnitude than the classical correlations. For this channel as well as 
the bit-phase-flip channel, appropriate quantities associated with the dynamics of the correlations 
can be derived which signal the occurrence of a QPT.
\end{abstract}


\maketitle

\section{Introduction}

The correlations which exist between the different constituents of an interacting quantum system 
have two distinct components: classical and quantum. The most well-known example of quantum 
correlations is that of entanglement which serves as a fundamental resource in several quantum
information processing tasks \cite{amico,lewenstein,horodecki}. In the case of bipartite quantum 
systems, the quantum discord (QD) has been proposed to quantify quantum correlations more general 
than those captured by entanglement \cite{ollivier,henderson,zurek}. In fact, there are separable
mixed states which by definition are unentangled but have non-zero QD. The utility of such states 
in certain computational tasks has recently been demonstrated both theoretically \cite{datta} 
and experimentally \cite{lanyon}. QD thus has the potential to serve as an important resource in 
certain types of quantum information processing tasks.

The quantum mutual information $I\left(\rho_{AB}\right)$ measures the total correlations, with 
classical as well as quantum components, in a bipartite quantum system and is given by 
\begin{eqnarray}
 I\left(\rho_{AB}\right)=S\left(\rho_{A}\right)+S\left(\rho_{B}\right)-S\left(\rho_{AB}\right)
 \label{mutual}
\end{eqnarray}
where $\rho_{AB}$ is the density matrix of the full system and $\rho_{A}\;\left(\rho_{B}\right)$ the 
reduced density matrix of subsystem A (B). Also, $S(\rho)$ represents the von Neumann entropy with
$S\left(\rho\right)=-\mbox{Tr}\left\{\rho \log_{2} \rho\right\}$. The QD 
, $Q\left(\rho_{AB}\right)$, is defined to be the difference between $I\left(\rho_{AB}\right)$ and
the classical correlations $C\left(\rho_{AB}\right)$, i.e., 
\begin{eqnarray}
 Q\left(\rho_{AB}\right)=I\left(\rho_{AB}\right)-C\left(\rho_{AB}\right)
 \label{qcorelation}
\end{eqnarray}
The computation of classical correlations, $C\left(\rho_{AB}\right)$, is carried out in the following 
manner \cite{henderson,luo,sarandy}. In classical information theory, the mutual information 
$I\left(A,B\right)=H(A)+H(B)-H(A,B)$ quantifies the total correlation between two random variables 
$A$ and $B$. $H(A)=-\sum_{a}p_{a}\log_{2}p_{a}$, $H(B)=-\sum_{b}p_{b}\log_{2}p_{b}$ and $H(A,B)
=-\sum_{a,b}p_{a,b}\log_{2}p_{a,b}$ are the Shannon entropies for the variable $A$, the variable $B$
and the joint system $AB$ respectively. The joint probability of the variables $A$ and $B$ having 
the values $a$ and $b$ respectively is represented by $p_{a,b}$ and $p_{a}=\sum_{b}p_{a,b}$, 
$p_{b}=\sum_{a}p_{a,b}$. The classical mutual information has an equivalent 
expression $J(A,B)=H(A)-H(A|B)$ via the Bayes\textquoteright rule. The conditional entropy $H(A|B)$ is a measure of 
our ignorance about the state of $A$ when that of $B$ is known. In the case of a quantum system, the
von Neumann entropy replaces the Shannon entropy and the quantum generalization of the classical 
mutual information $I(A,B)$ is straightforward yielding the expression in equation (\ref{mutual}). The
quantum version of $J\left(A,B\right)$ is not, however, equivalent to $I\left(\rho_{AB}\right)$. 
This is because the magnitude of the quantum conditional entropy depends on the type of measurement 
performed on subsystem $B$ to gain knowledge of its state so that different measurement choices yield 
different results. We consider von Neumann-type measurements on $B$ defined in terms of a complete 
set of orthogonal projectors, ${\Pi_{i}}$, corresponding to the set of possible outcomes $i$. The
state of the system, once the measurement is made, is given by 
\begin{eqnarray}
 \rho_{i}=\left(I\otimes\Pi_{i}^{B}\right)\rho_{AB}\left(I\otimes\Pi_{i}^{B}\right)/p_{i}
 \label{state}
\end{eqnarray}
with 
\begin{eqnarray}
 p_{i}=\mbox{Tr}\left(\left(I\otimes\Pi_{i}^{B}\right)\rho_{AB}\left(I\otimes\Pi_{i}^{B}\right)\right)
 \label{prob}
\end{eqnarray}
Here $I$ denotes the identity operator for the subsystem $A$ and $p_{i}$ gives the probability of
obtaining the outcome $i$. The quantum analogue of the conditional entropy is
\begin{eqnarray}
 S\left(\rho_{AB}|\left\{\Pi_{i}^{B}\right\}\right)=\sum_{i}p_{i}S\left(\rho_{i}\right)
 \label{conditional}
\end{eqnarray}
The quantum extension of the classical mutual information is given by 
\begin{eqnarray}
 J\left(\rho_{AB}|\left\{\Pi_{i}^{B}\right\}\right)=S\left(\rho_{A}\right)-S\left(\rho_{AB}|\left\{\Pi_{i}^{B}\right\}\right)
 \label{qmutual}
\end{eqnarray}
When projective measurements are made on the subsystem $B$, the non-classical correlations between the 
subsystems are removed. Since the value of $J\left(\rho_{AB}|\left\{\Pi_{i}^{B}\right\}\right)$ 
is dependent on the choice of $\left\{\Pi_{i}\right\}$, $J$ should be maximized over all 
$\left\{\Pi_{i}\right\}$ to ensure that it contains the whole of the classical correlations. 
Thus the quantity 
\begin{eqnarray}
 C\left(\rho_{AB}\right)=\underset{\left\{\Pi_{i}^{B}\right\}}{\mbox{max}}
                         \left(J\left(\rho_{AB}|\left\{\Pi_{i}^{B}\right\}\right)\right)
 \label{classical}
\end{eqnarray}
provides a quantitative measure of the total classical correlations \cite{henderson}.

Though the concept of the QD is firmly established, its computation is restricted to two-qubit 
states and that too when $\rho_{AB}$ has special forms \cite{luo,sarandy,ali}. For two-qubit 
$X$-states, the density matrix in the basis 
$\left\{|11\rangle,|10\rangle,|01\rangle,|00\rangle\right\}$ has the general structure 
\begin{eqnarray}
\rho_{X}= \left(
 \begin{array}{cccc}
 \rho_{11} & 0 & 0 & \rho_{14} \\
 0 & \rho_{22} & \rho_{23} & 0 \\
 0 & \rho_{32} & \rho_{33} & 0 \\
 \rho_{41} & 0 & 0 & \rho_{44} 
 \end{array}
 \right)
 \label{DEN}
\end{eqnarray}
with $\rho_{12}=\rho_{21}=\rho_{13}=\rho_{31}=\rho_{24}=\rho_{42}=\rho_{34}=\rho_{43}=0$. Analytic 
expressions for the QD can be derived only in some special cases. We restrict attention to the case 
\begin{eqnarray}
\rho_{AB}= \left(
 \begin{array}{cccc}
 a & 0 & 0 & f \\
 0 & b & z & 0 \\
 0 & z & b & 0 \\
 f & 0 & 0 & d 
 \end{array}
 \right)
 \label{DEN2}
\end{eqnarray}
where $A$, $B$ correspond to the two individual qubits and $z$, $f$ are real numbers. The eigenvalues 
of $\rho_{AB}$ are \cite{sarandy}
\begin{eqnarray}
 \lambda_{0}&=&\frac{1}{4}\left\{\left(1+c_{3}\right)+\sqrt{4 c_{4}^{2}+\left(c_{1}-c_{2}\right)^{2}}\right\} \nonumber \\
 \lambda_{1}&=&\frac{1}{4}\left\{\left(1+c_{3}\right)-\sqrt{4 c_{4}^{2}+\left(c_{1}-c_{2}\right)^{2}}\right\} \nonumber \\
 \lambda_{2}&=&\frac{1}{4}\left(1-c_{3}+c_{1}+c_{2}\right) \nonumber \\
 \lambda_{3}&=&\frac{1}{4}\left(1-c_{3}-c_{1}-c_{2}\right)
 \label{eigenvalues}
\end{eqnarray}
with
\begin{eqnarray}
 c_{1}&=&2z+2f \nonumber \\
 c_{2}&=&2z-2f \nonumber \\
 c_{3}&=&a+d-2b \nonumber \\
 c_{4}&=&a-d 
 \label{cvalues}
\end{eqnarray}
The mutual information $I\left(\rho_{AB}\right)$ (equation (\ref{mutual})) can be written as \cite{luo,sarandy}
\begin{eqnarray}
 I\left(\rho_{AB}\right)=S\left(\rho_{A}\right)+S\left(\rho_{B}\right)
                         +\sum_{\alpha=0}^{3}\lambda_{\alpha}\log_{2}\lambda_{\alpha}
\label{ivalue}
\end{eqnarray}
where 
\begin{eqnarray}
 S\left(\rho_{A}\right)=S\left(\rho_{B}\right)&=&-\frac{1+c_{4}}{2}\log_{2}\frac{1+c_{4}}{2} \nonumber \\
 &&-\frac{1-c_{4}}{2}\log_{2}\frac{1-c_{4}}{2}
\label{redvalue}                                               
\end{eqnarray}
With the expressions for $I\left(\rho_{AB}\right)$ and $C\left(\rho_{AB}\right)$ given in 
equations (\ref{ivalue}), (\ref{redvalue}) and (\ref{classical}) respectively, the QD, 
$Q\left(\rho_{AB}\right)$, (equation (\ref{qcorelation})) can in principle be computed. The difficulty 
lies in the maximization procedure to be carried out in order to compute $C\left(\rho_{AB}\right)$.
When $\rho_{AB}$ is of the form given in (\ref{DEN2}), the maximization can be done analytically
\cite{fanchini} resulting in the following expression for the QD:
\begin{eqnarray}
 Q\left(\rho_{AB}\right)=\mbox{min}\left\{Q_{1},Q_{2}\right\}
\label{discord}
\end{eqnarray}
where 
\begin{eqnarray}
 Q_{1}&=&S\left(\rho_{B}\right)-S\left(\rho_{AB}\right)-a\log_{2}\frac{a}{a+b}-b\log_{2}\frac{b}{a+b}\nonumber \\
 &&-d\log_{2}\frac{d}{d+b}-b\log_{2}\frac{b}{d+b}
 \label{discord1}
\end{eqnarray}
and
\begin{eqnarray}
 Q_{2}&=&S\left(\rho_{B}\right)-S\left(\rho_{AB}\right)-\Delta_{+}\log_{2}\Delta_{+}\nonumber \\
      &&-\Delta_{-}\log_{2}\Delta_{-}
 \label{discord2}
\end{eqnarray}
with $\Delta_{\pm}=\frac{1}{2}\left(1\pm\Gamma \right)$ and 
$\Gamma^{2}=\left(a-d\right)^{2}+4\left(|z|+|f|\right)^{2}$
   
Quantum systems, in general, are open systems because of the inevitable interaction of a system with 
its environment. This results in decoherence, i.e., a gradual loss from a coherent superposition to 
a statistical mixture with an accompanying decay of the quantum correlations in composite systems. 
The dynamics of entanglement and QD under system-environment interactions have been investigated in a 
number of recent studies \cite{maziero1,almeida,werlang1,maziero2,mazzola,pal}. One feature which 
emerges out of such studies is that the QD is more robust than entanglement in the case of Markovian
(memoryless) dynamics. The dynamics may bring about the complete disappearance of 
entanglement at a finite time termed the \textquoteleft entanglement sudden death\textquoteright 
\cite{maziero1,almeida}. The QD, however, is found to decay in time but vanishes only asymptotically
\cite{werlang1,mazzola,pal,ferraro}. Also, under Markovian time evolution and for a class of states, the 
decay rates of the classical and quantum correlations exhibit sudden changes in behaviour 
\cite{maziero2,mazzola}. Three general types of dynamics under the effect of decoherence have been 
observed \cite{maziero2}: (i) $C\left(\rho_{AB}\right)$ is constant and $Q\left(\rho_{AB}\right)$
decays monotonically as a function of time, (ii) $C\left(\rho_{AB}\right)$ decays monotonically over 
time till a parametrized time $p_{sc}$ and remains constant thereafter. $Q\left(\rho_{AB}\right)$
has an abrupt change in the decay rate at $p_{sc}$ and has magnitude greater than that of 
$C\left(\rho_{AB}\right)$ in a parametrized time interval and (iii) both $C\left(\rho_{AB}\right)$
and $Q\left(\rho_{AB}\right)$ decay monotonically. Mazzola et al. \cite{mazzola} have
demonstrated that under Markovian dynamics (qubits interacting with non-dissipative independent 
reservoirs) and for a class of initial states the QD remains constant in a finite time interval 
$0<t<\tilde{t}$. In this time interval, the classical correlations, $C\left(\rho_{AB}\right)$, 
decay monotonically. Beyond $t=\tilde{t}$, $C\left(\rho_{AB}\right)$ becomes constant while the QD
decreases monotonically with time. The sudden change in the decay rate of correlations and their 
constancy in certain time intervals have been demonstrated in actual experiments \cite{xu,auccaise}.

In this paper, we focus on a two-qubit system each qubit of which interacts with an independent 
reservoir. The density matrix of the two-qubit system is described by the reduced density matrix 
derived from the ground state density matrix of the transverse Ising model (TIM) in one dimension (1d).
We investigate the dynamics of the QD as well as the classical correlations under Markovian time 
evolution and identify some new features close to the quantum critical point of the model.
In Sec. II, the calculational scheme for studying the dynamics of the classical and quantum
correlations is introduced. We further describe the quantum channels representing the 
system-environment interactions for which the computations are carried out. Sec. III presents 
the major results obtained and a discussion thereof. Sec. IV contains some concluding remarks.

\section{Dynamics of classical and quantum correlations}
We consider the TIM Hamiltonian in 1d described by the Hamiltonian
\begin{eqnarray}
 H=-\lambda\sum_{j=1}^{L}\sigma_{j}^{x}\sigma_{j+1}^{x}-\sum_{j=1}^{L}\sigma_{j}^{z}
\label{TIM}
\end{eqnarray}
where $\sigma_{j}^{x}$ and $\sigma_{j}^{z}$ are the Pauli matrices defined at the site $j$ of the 
chain and $L$ is the total number of sites in the chain. We further assume periodic boundary conditions.
The Hamiltonian in (\ref{TIM}) can be exactly diagonalized in the thermodynamic limit 
$L\rightarrow\infty$ \cite{dutta,osborne}. When the parameter $\lambda=0$, all the spins are oriented
in the positive $z$ direction in the ground state whereas for $\lambda=\infty$, the ground state is 
doubly degenerate with all the spin pointing in either the positive or the negative direction. 
As one goes from one limit to the other, a quantum phase transition 
(QPT) occurs at the critical point $\lambda_{c}=1$ separating two different phases, the paramagnetic 
phase with the magnetization $\langle\sigma^{x}\rangle$ zero and the ordered ferromagnetic phase 
characterized by a non-zero magnetization. The QPT is signaled by the divergence of the correlation 
length at the critical point. Since the ground state wave function undergoes a qualitative change 
at the critical point, it is reasonable to expect that the quantum correlations present in the ground 
state would provide signatures of the occurrence of a QPT. Such signatures in fact do exist for different 
measures of quantum correlations, namely, entanglement \cite{amico,lewenstein,dutta,osborne,osterloh}
and QD \cite{sarandy,dillenschneider,werlang2}. For the TIM, the two-site reduced density matrix 
$\rho_{ij}$ has the form given in equation (\ref{DEN2}) \cite{dutta,osborne,dillenschneider,syljuasen} 
with
\begin{eqnarray}
 a&=&\frac{1}{4}+\frac{\langle\sigma^{z}\rangle}{2}+
     \frac{\langle\sigma_{i}^{z}\sigma_{j}^{z}\rangle}{4} \nonumber \\
 d&=&\frac{1}{4}-\frac{\langle\sigma^{z}\rangle}{2}+
     \frac{\langle\sigma_{i}^{z}\sigma_{j}^{z}\rangle}{4} \nonumber \\
 b&=&\frac{1}{4}\left(1-\langle\sigma_{i}^{z}\sigma_{j}^{z}\rangle\right) \nonumber \\
 z&=&\frac{1}{4}
     \left(\langle\sigma_{i}^{x}\sigma_{j}^{x}\rangle+\langle\sigma_{i}^{y}\sigma_{j}^{y}\rangle\right) \nonumber \\
 f&=&\frac{1}{4}
     \left(\langle\sigma_{i}^{x}\sigma_{j}^{x}\rangle-\langle\sigma_{i}^{y}\sigma_{j}^{y}\rangle\right) 
 \label{elements}
\end{eqnarray}
The magnetization $\langle\sigma^{z}\rangle$ of the TIM is given by \cite{dutta,dillenschneider} 
\begin{eqnarray}
 \langle\sigma^{z}\rangle=-\frac{1}{\pi}\int_{0}^{\pi}d\phi\frac{\left(1+\lambda\cos{\phi}\right)}
                          {\omega_{\phi}}
\end{eqnarray}
where 
\begin{eqnarray}
 \omega_{\phi}=\sqrt{\left(\lambda\sin{\phi}\right)^{2}+\left(1+\lambda\cos{\phi}\right)^{2}}
\end{eqnarray}
is the energy spectrum. The spin-spin correlation functions are obtained from the determinant of 
Toeplitz matrices \cite{dillenschneider,barouch,pfeuty}
\begin{eqnarray}
  \left\langle\sigma_{i}^{x}\sigma_{i+r}^{x}\right\rangle &=& 
 \left|
 \begin{array}{cccc}
  G_{-1} & G_{-2} & \cdots & G_{-r}\\
  G_{0} & G_{-1} & \cdots & G_{-r+1}\\
  \vdots & \vdots & \ddots & \vdots\\
  G_{r-2} & G_{r-3} & \cdots & G_{-1}\\
 \end{array}\right| \nonumber \\
  \left\langle\sigma_{i}^{y}\sigma_{i+r}^{y}\right\rangle &=& 
 \left|
 \begin{array}{cccc}
  G_{1} & G_{0} & \cdots & G_{-r+2}\\
  G_{2} & G_{1} & \cdots & G_{-r+3}\\
  \vdots & \vdots & \ddots & \vdots\\
  G_{r} & G_{r-1} & \cdots & G_{1}\\
 \end{array}\right| \nonumber \\
 \left\langle\sigma_{i}^{z}\sigma_{i+r}^{z}\right\rangle &=&
 \left\langle\sigma^{z}\right\rangle^{2}-G_{r}G_{-r} 
\end{eqnarray}
where 
\begin{eqnarray}
  G_{r}&=&\frac{1}{\pi}\int_{0}^{\pi}d\phi\cos(r\phi)
 \frac{\left(1+\lambda\cos\phi\right)}{\omega_{\phi}}\nonumber \\
 &&-\frac{\lambda}{\pi}\int_{0}^{\pi}d\phi\sin(r\phi)\frac{\sin\phi}{\omega_{\phi}}
 \label{spincor}
\end{eqnarray}

\begin{figure}
\begin{center}
\includegraphics[scale=0.6]{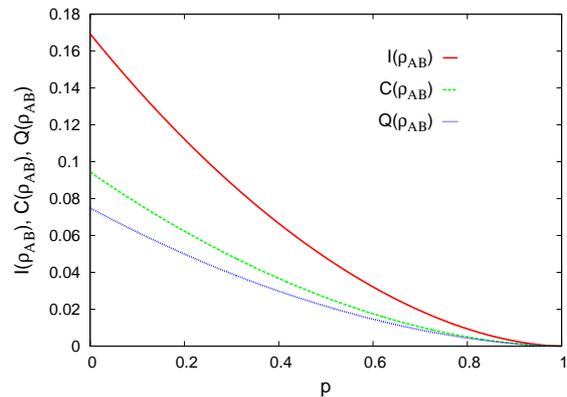}
\par\end{center}
\caption{(Color online) Amplitude damping channel: decay of mutual information $I\left(\rho_{AB}\right)$
(solid line), classical correlations $C\left(\rho_{AB}\right)$ (dashed line) and quantum correlations (QD) 
$Q\left(\rho_{AB}\right)$ (dotted line) as a function of parametrized time $p=1-e^{-\gamma t}$ and 
$\lambda=0.5$}
\label{ad}
\end{figure}

We next consider the interaction of the chain of qubits, each qubit representing an Ising spin, with 
an environment. We choose the initial state of the whole system at time $t=0$ to be of the product 
form, i.e., 
\begin{eqnarray}
 \rho(0)=\rho_{s}(0)\otimes\rho_{e}(0)
 \label{sysden}
\end{eqnarray}
where the density matrices $\rho_{s}$ and $\rho_{e}$ correspond to the system and environment 
respectively. We assume that the environment is represented by $L$ independent reservoirs each of 
which interacts locally with a qubit constituting the Ising chain. The two-qubit reduced density matrix 
obtained from equation (\ref{sysden}) can be written as
\begin{eqnarray}
 \rho_{r}(0)=\rho_{rs}(0)\otimes\rho_{re}(0)
\end{eqnarray}
where $\rho_{rs}$ and $\rho_{re}$ represent the two-qubit reduced density matrix of the transverse 
Ising chain and the corresponding reduced density matrix of the two-reservoir environment 
respectively. The two-qubit reduced density matrix $\rho_{rs}$ is obtained by taking partial 
trace on $\rho_{s}$ over the states of all the qubits other than the two chosen qubits. Similarly, 
$\rho_{re}$ is obtained from $\rho_{e}$ by taking a partial trace over the states of all the 
reservoirs other than the two local reservoirs of the selected qubits. 
The quantum channel describing the interaction between a qubit and its environment can 
be of various types: amplitude damping, phase damping, bit-flip, phase-flip, bit-phase-flip etc. 
\cite{maziero1,nielsenchuang}. Our objective is to investigate the dynamics of the two-qubit classical
and quantum correlations (in the form of the QD) under the influence of various quantum channels.

The time evolution of the closed quantum system, comprised of both the system and the environment, 
is given by 
\begin{eqnarray}
 \rho_{se}(t)=U(t)\rho_{se}(0)U^{\dagger}(t)
 \label{evolve2}
\end{eqnarray}
where $U(t)$ is the unitary evolution operator generated by the total Hamiltonian $H$ of the 
system $\left(U=e^{-iHt/\hbar}\right)$. $H$ is given by $H=H_{s}+H_{e}+H_{se}$ where $H_{s}$ 
and $H_{e}$ represent the bare system and environment Hamiltonians respectively and $H_{se}$ 
the Hamiltonian describing the interactions between the system and the environment. The time 
evolution of the system $s$ subject to the influence of the environment $e$ is obtained by 
carrying out a partial trace on $\rho_{se}(t)$ (equation (\ref{evolve2})) over the environment 
states, i.e.,
\begin{eqnarray}
 \rho_{s}(t)=\mbox{Tr}_{e}\left[U(t)\rho_{se}(0)U^{\dagger}(t)\right]
\end{eqnarray}
Let $|e_{k}\rangle$ be an orthogonal basis spanning the finite-dimensional state space of the 
environment. With the initial state of the whole system given by equation (\ref{sysden}),
\begin{eqnarray}
 \rho_{s}(t)=\sum_{k}\langle e_{k}|U\left[\rho_{s}(0)\otimes\rho_{e}(0)\right]U^{\dagger}|e_{k}\rangle
\end{eqnarray}
Let $\rho_{e}(0)=|e_{0}\rangle\langle e_{0}|$ be the initial state of the environment. Then
\begin{eqnarray}
 \rho_{s}(t)=\sum_{k}E_{k}\rho_{s}(0) E_{k}^{\dagger}
 \label{sysev}
\end{eqnarray}
where $E_{k}\equiv\langle e_{k}|U|e_{0}\rangle$ is the Kraus operator which acts on the state space of 
the system only \cite{maziero1,nielsenchuang}. Let $\left\{\phi_{i}\right\}$, $i=1,2,...,d,$ define 
the basis in the state space of the system $s$. There are then at most $d^{2}$ independent 
Kraus operators $E_{k}$, $k=0,...,d^{2}-1$ \cite{nielsenchuang,salles}. The unitary evolution of 
$s+e$ is given by the map:
\begin{eqnarray}
 |\phi_{1}\rangle|e_{0}\rangle &\rightarrow& E_{0}|\phi_{1}\rangle|e_{0}\rangle+...
                                             +E_{d^{2}-1}|\phi_{1}\rangle|e_{d^{2}-1}\rangle \nonumber \\
 |\phi_{2}\rangle|e_{0}\rangle &\rightarrow& E_{0}|\phi_{2}\rangle|e_{0}\rangle+...
                                             +E_{d^{2}-1}|\phi_{2}\rangle|e_{d^{2}-1}\rangle \nonumber \\
 &\vdots& \nonumber \\
 |\phi_{d}\rangle|e_{0}\rangle &\rightarrow& E_{0}|\phi_{d}\rangle|e_{0}\rangle+...
                                             +E_{d^{2}-1}|\phi_{d}\rangle|e_{d^{2}-1}\rangle 
\label{map1}
\end{eqnarray}
In compact notation, the map is given by 
\begin{eqnarray}
 U|\phi_{i}\rangle|e_{0}\rangle\equiv\sum_{k}E_{k}|\phi_{i}\rangle|e_{k}\rangle, \; i=1,2,...,d
\label{map2}
\end{eqnarray}
In the case of $N$ system parts with each part interacting with a local independent environment, 
equation (\ref{sysev}) becomes
\begin{eqnarray}
 \rho_{s}(t)=\sum_{k_{1},..,k_{N}}E_{k_{1}}^{(1)}\otimes..\otimes E_{k_{N}}^{(N)}\rho_{s}(0)
             E_{k_{1}}^{(1)\dagger}\otimes..\otimes E_{k_{N}}^{(N)\dagger}\nonumber \\
 \;\;\;
\label{npart}
\end{eqnarray}
where $E_{k_{\alpha}}^{(\alpha)}$ is the $k_{\alpha}$th Kraus operator with the environment acting on system part 
$\alpha$. The specific form for $\rho_{s}(t)$ arises as the total evolution operator can be written 
as $U(t)=U_{1}(t)\otimes U_{2}(t) \otimes ... \otimes U_{N}(t)$.
Following the general formalism of the Kraus operator representation, an initial state, 
$\rho_{rs}(0)$, of the two-qubit reduced density matrix evolves as \cite{werlang1,nielsenchuang}
\begin{eqnarray}
 \rho_{rs}(t)=\sum_{\mu,\nu}E_{\mu,\nu}\rho_{rs}(0)E_{\mu,\nu}^{\dagger}
\label{evolve}
\end{eqnarray}
where the Kraus operators $E_{\mu,\nu}=E_{\mu}\otimes E_{\nu}$ satisfy the completeness relation
$\sum_{\mu,\nu}E_{\mu,\nu}E_{\mu,\nu}^{\dagger}=I$ for all $t$. We now briefly describe the various 
quantum channels considered in the paper and write down the corresponding Kraus operators. A fuller 
description can be obtained from Refs. \cite{werlang1,nielsenchuang}.

\begin{figure}
\begin{center}
\includegraphics[scale=0.6]{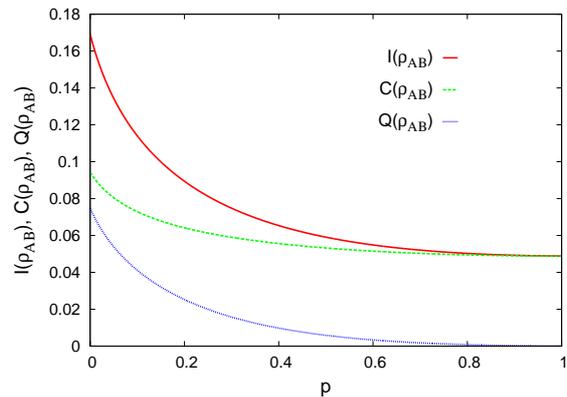}
\par\end{center}
\caption{(Color online) Bit-flip channel: decay of mutual information $I\left(\rho_{AB}\right)$
(solid line), classical correlations $C\left(\rho_{AB}\right)$ (dashed line) and quantum correlations (QD) 
$Q\left(\rho_{AB}\right)$ (dotted line) as a function of parametrized time $p=1-e^{-\gamma t}$ and 
$\lambda=0.5$}
\label{bf}
\end{figure}

\noindent (i) \textit{Amplitude Damping Channel.}

The channel describes the dissipative interaction between a system and its environment resulting in 
an exchange of energy between $s$ and $e$ so that $s$ is ultimately in thermal equilibrium with 
$e$. The  $s+e$ time evolution is given by the unitary transformation
\begin{eqnarray}
 |0\rangle_{s}|0\rangle_{e} &\rightarrow& |0\rangle_{s}|0\rangle_{e}
\label{map31}
\end{eqnarray}
\begin{eqnarray}
 |1\rangle_{s}|0\rangle_{e} &\rightarrow& \sqrt{q}|1\rangle_{s}|0\rangle_{e}+\sqrt{p}|0\rangle_{s}|1\rangle_{e}
 \label{map32}
\end{eqnarray}
where $|0\rangle_{s}$ and $|1\rangle_{s}$ are the ground and excited qubit states and $|0\rangle_{e}$,
$|1\rangle_{e}$ denote states of the environment with no excitation (vacuum state) and one excitation
respectively. Equation (\ref{map31}) stipulates that there is no dynamic evolution if the system 
and the environment are in their ground states. Equation (\ref{map32}) states that if the system qubit 
is in the excited state, the probability to remain in the same state is $q$ and the probability 
of decaying to the ground state is $p$ $(p+q=1)$. The decay of the qubit state is accompanied by 
a transition of the environment to a state with one excitation. The qubit states may be two atomic 
states with the excited state decaying to the ground state by emitting a photon. The environment
on acquiring the photon is no longer in the vacuum state. With a knowledge of the map equations 
(equations (\ref{map31}) and (\ref{map32})), the Kraus operators for the amplitude damping channel 
can be written as
\begin{eqnarray}
 E_{0}=\left(
 \begin{array}{cc}
  1 & 0 \\
  0 & \sqrt{q}
 \end{array}\right);\;\;
 E_{1}=\left(
 \begin{array}{cc}
  0 & \sqrt{p} \\
  0 & 0
 \end{array}\right)
\end{eqnarray}
where $q=1-p$. The Kraus operators for the two distinct environments (one for each qubit) have 
identical forms. In the case of Markovian time evolution, $p$ is given by  $p=1-e^{-\gamma t}$ 
with $\gamma$ denoting the decay rate.    

\noindent (ii) \textit{Phase Damping (dephasing) Channel.}
The channel describes the loss of quantum coherence without loss of energy. The Kraus operators are:
\begin{eqnarray}
 E_{0}=\left(
 \begin{array}{cc}
  1 & 0 \\
  0 & \sqrt{q}
 \end{array}\right);\;\;
 E_{1}=\left(
 \begin{array}{cc}
  0 & 0 \\
  0 & \sqrt{p}
 \end{array}\right)
\end{eqnarray}  
with $q=1-p$ and $p=1-e^{-\gamma t}$.

\begin{figure}
\begin{center}
  \includegraphics[scale=0.6]{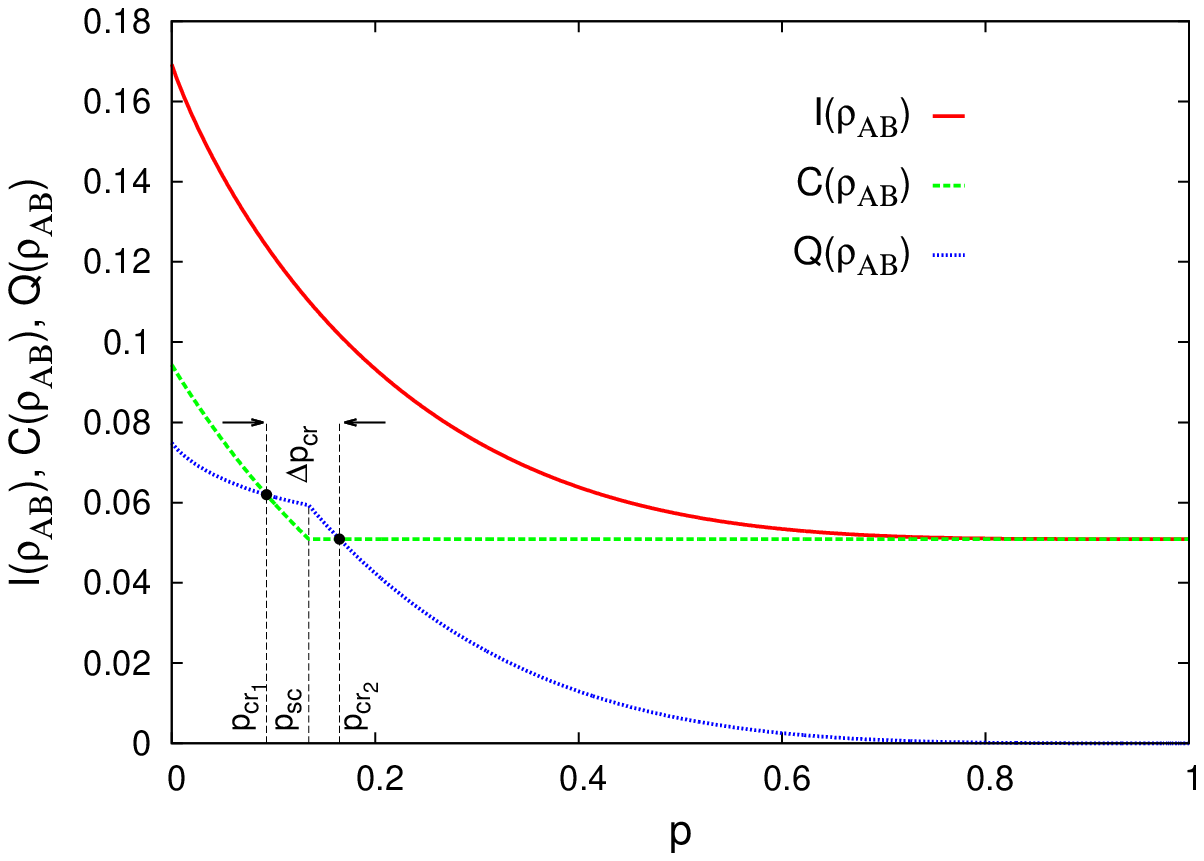}
\par\end{center}
\caption{(Color online) Phase-flip channel: decay of mutual information $I\left(\rho_{AB}\right)$
(solid line), classical correlations $C\left(\rho_{AB}\right)$ (dashed line) and quantum correlations (QD) 
$Q\left(\rho_{AB}\right)$ (dotted line) as a function of parametrized time $p=1-e^{-\gamma t}$ and 
$\lambda=0.5$. Also, $p_{cr_{1}}=0.0932$, $p_{sc}=0.1347$ and $p_{cr_{2}}=0.1649$}
\label{pf1}
\end{figure}

\noindent (iii) \textit{Bit-flip, phase-flip and bit-phase-flip channels.}      
The channels destroy the information contained in the phase relations without involving an exchange 
of energy. The Kraus operators are
\begin{eqnarray}
 E_{0}=\sqrt{q^{\prime}}\left(
 \begin{array}{cc}
  1 & 0 \\
  0 & 1
 \end{array}\right);\;\;
 E_{1}=\sqrt{p/2}\sigma_{i}
\end{eqnarray}  
where $i=x$ for the bit-flip, $i=y$ for the bit-phase-flip and $i=z$ for the phase-flip channel with
$q^{\prime}=1-p/2$ and  $p=1-e^{-\gamma t}$. The expanded forms of the Kraus operators are:

\noindent Bit-flip
\begin{eqnarray}
  E_{0}&=&\left(
 \begin{array}{cc}
  \sqrt{1-p/2} & 0 \\
  0 & \sqrt{1-p/2}
 \end{array}\right) \nonumber \\
 E_{1}&=&\left(
 \begin{array}{cc}
  0 & \sqrt{p/2} \\
  \sqrt{p/2} & 0
 \end{array}\right)
\end{eqnarray}
\noindent Phase-flip
\begin{eqnarray}
  E_{0}&=&\left(
 \begin{array}{cc}
  \sqrt{1-p/2} & 0 \\
  0 & \sqrt{1-p/2}
 \end{array}\right)\nonumber \\
 E_{1}&=&\left(
 \begin{array}{cc}
  \sqrt{p/2} & 0 \\
  0 & -\sqrt{p/2}
 \end{array}\right)
\end{eqnarray}
\noindent Bit-phase-flip
\begin{eqnarray}
  E_{0}&=&\left(
 \begin{array}{cc}
  \sqrt{1-p/2} & 0 \\
  0 & \sqrt{1-p/2}
 \end{array}\right) \nonumber \\
 E_{1}&=&\left(
 \begin{array}{cc}
  0 & -i\sqrt{p/2} \\
  i\sqrt{p/2} & 0
 \end{array}\right)
\end{eqnarray}
As shown in Ref. \cite{nielsenchuang}, the phase damping quantum operation is identical to that of 
the phase-flip channel so that we will consider only one of these, the phase-flip channel, in the 
following. 

For a specific quantum channel, it is now straightforward to calculate the dynamics of the classical
and quantum correlations. Equation (\ref{evolve}) describes the time evolution of the reduced density 
matrix of the TIM subjected to the influence of an environment via a quantum channel. The initial
state $\rho_{rs}(0)$ has the form given in equation (\ref{DEN2}) the elements of which are known via 
the equations (\ref{elements})-(\ref{spincor}). The time-evolved state $\rho_{rs}(t)$ has again the form
given in equation (\ref{DEN2}) with the time dependence occurring in only the off-diagonal elements. With 
a knowledge of the elements, the mutual information $I(\rho_{AB})$, the QD $Q(\rho_{AB})$ and the 
classical correlations $C(\rho_{AB})$ can be computed at any time $t$ with the help of the formulae 
in equations (\ref{eigenvalues})-(\ref{discord2}) and equation (\ref{qcorelation}). The results of our 
calculations for the various quantum channels are described in the next Section.

\section{Results}

\begin{figure}
\begin{center}
  \includegraphics[scale=0.6]{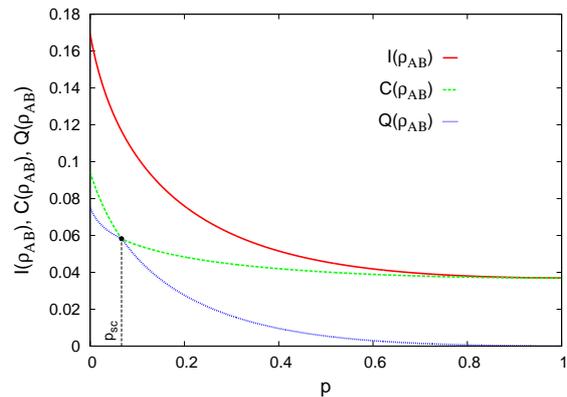}
\par\end{center}
\caption{(Color online) Bit-phase-flip channel: decay of mutual information $I\left(\rho_{AB}\right)$
(solid line), classical correlations $C\left(\rho_{AB}\right)$ (dashed line) and quantum correlations (QD) 
$Q\left(\rho_{AB}\right)$ (dotted line) as a function of parametrized time $p=1-e^{-\gamma t}$ and 
$\lambda=0.5$. Also, $p_{sc}=0.0666$}
\label{bpf1}
\end{figure}

In the following, we make the substitution $\rho_{AB}=\rho_{rs}(t)$.

\noindent \textit{Amplitude Damping Channel.}
The dynamical evolution of the mutual information $I(\rho_{AB})$, the classical correlations 
$C(\rho_{AB})$ and the QD $Q(\rho_{AB})$ as a function of the parametrized time $p$ 
$\left(p=1-e^{-\gamma t}\right)$ is shown in Fig.\ref{ad} for $\lambda=0.5$. The solid, dashed 
and dotted lines represent the variations of $I(\rho_{AB})$, $C(\rho_{AB})$ and $Q(\rho_{AB})$ 
respectively with $p$. All the correlations decay to zero in the asymptotic limit of 
$t\rightarrow\infty$, i.e., $p\rightarrow 1$. There is further no parametrized time interval or point 
when the quantum correlation becomes greater than the classical correlation. 

\noindent \textit{Bit-flip Channel.}
Fig.\ref{bf} exhibits the dynamical evolution of $I(\rho_{AB})$ (solid line), 
$C(\rho_{AB})$ (dashed line) and $Q(\rho_{AB})$ (dotted line) as a function of the parametrized 
times and with $\lambda=0.5$. The quantum correlations disappear completely in the asymptotic limit 
$p\rightarrow 1$. In the same limit,  $I(\rho_{AB})=C(\rho_{AB})$ has a finite value. In the case of 
both the amplitude damping and bit-flip channels, the same features as observed for 
$\lambda=0.5$ are obtained for the other values of $\lambda$.

\begin{figure}
\begin{center}
  \includegraphics[scale=0.6]{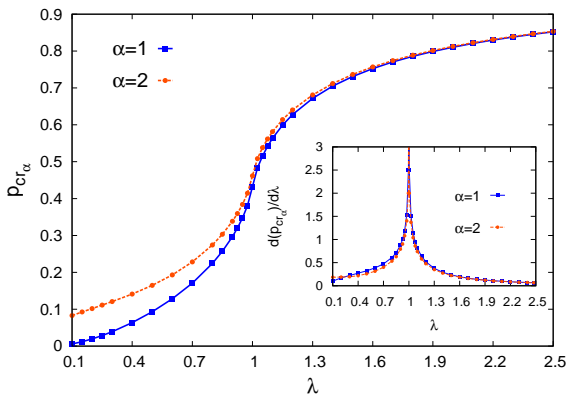}
\par\end{center}
\caption{(Color online) Variations of $p_{cr_{1}}$ (solid line) and $p_{cr_{2}}$ (dashed line) 
with $\lambda$ in the case of the phase-flip channel; (inset) The first derivative of $p_{cr_{1}}$
and $p_{cr_{2}}$ w.r.t $\lambda$ diverges as the QCP $\lambda_{c}=1$ is approached }
\label{pf2}
\end{figure}

\noindent \textit{Phase-flip Channel.}     
In Fig.\ref{pf1}, we plot the variations of $I\left(\rho_{AB}\right)$ (solid line), 
$C\left(\rho_{AB}\right)$ (dashed line) and $Q\left(\rho_{AB}\right)$ (dotted line) versus the 
parametrized time $p$ with $\lambda=0.5$. There is a sudden change in the decay rates of both 
$C\left(\rho_{AB}\right)$ and $Q\left(\rho_{AB}\right)$ at $p=p_{sc}$. There are two points, 
$p=p_{cr_{1}}$ and $p=p_{cr_{2}}$ at which the plots of $C\left(\rho_{AB}\right)$ and $Q\left(\rho_{AB}\right)$
cross each other. 
The classical correlations remains constant beyond the point $p=p_{sc}$ whereas the QD decays 
asymptotically to zero. In the parametrized time interval $p_{sc}\leq p\leq 1$, the magnitude 
of $C\left(\rho_{AB}\right)=\left.I\left(\rho_{AB}\right)\right|_{p=1}$, the mutual information 
of the completely decohered state $\left(p=1\right)$. In the interval $p_{cr_{1}}< p < p_{cr_{2}}$,
the quantum correlations are larger in magnitude than the classical correlations contradicting 
an earlier conjecture that $C\left(\rho_{AB}\right)\geq Q\left(\rho_{AB}\right)$ in any quantum 
state \cite{maziero2}. At the crossing points, $p_{cr_{1}}$ and $p_{cr_{2}}$, one gets the 
equality $C\left(\rho_{AB}\right)=Q\left(\rho_{AB}\right)=\frac{I\left(\rho_{AB}\right)}{2}$. 
Xu et al. \cite{xu} have recently investigated the dynamics of classical and quantum correlations under 
decoherence in an all-optical experimental setup. Fig.4 of their paper provides experimental 
verification of the dynamics displayed in Fig.\ref{pf1}.

\begin{figure}
\begin{center}
  \includegraphics[scale=0.6]{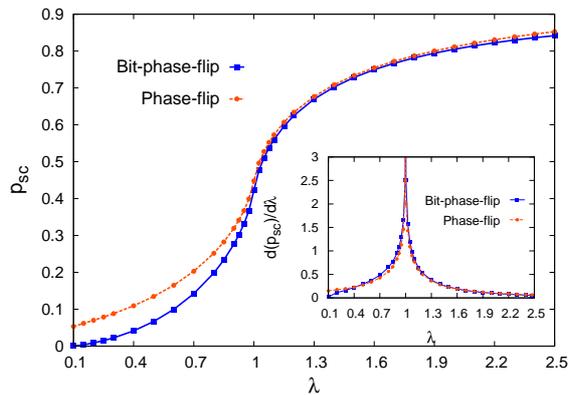}
\par\end{center}
\caption{(Color online) Variations of $p_{sc}$ with $\lambda$ for the phase-flip
(dashed line) and the bit-phase-flip (solid line) channels; (inset) The first derivative of $p_{sc}$
w.r.t $\lambda$ exhibits a divergence as the QCP $\lambda_{c}=1$ is approached}
\label{bpf2}
\end{figure}

The sudden change in the decay rates of $Q\left(\rho_{AB}\right)$ and $C\left(\rho_{AB}\right)$ 
at $p=p_{sc}$ is understood by noting that for $p<p_{sc}$, $Q=Q_{2}$ (equation (\ref{discord})) 
and for $p>p_{sc}$, $Q=Q_{1}$ with $C\left(\rho_{AB}\right)$ given by 
$C\left(\rho_{AB}\right)=I\left(\rho_{AB}\right)-Q\left(\rho_{AB}\right)$. At the crossing points 
$p_{cr_{1}}$ and $p_{cr_{2}}$, $Q\left(\rho_{AB}\right)=C\left(\rho_{AB}\right)$ so that 
$p_{cr_{1}}$ and $p_{cr_{2}}$ are the solutions of the equations 
$Q_{2}=\left(I\left(\rho_{AB}\right)-Q_{2}\right)$ and $Q_{1}=\left(I\left(\rho_{AB}\right)-Q_{1}\right)$  
respectively. The constancy of $C\left(\rho_{AB}\right)$ for values of $p>p_{sc}$ is explained by the 
fact that $Q=Q_{1}$ in this regime. From equations (\ref{mutual}), (\ref{qcorelation}) and 
(\ref{discord1}),
\begin{eqnarray}
 C\left(\rho_{AB}\right)&=&S\left(\rho_{A}\right)+a\log_{2}\frac{a}{a+b}+b\log_{2}\frac{b}{a+b}\nonumber \\
 &&+d\log_{2}\frac{d}{d+b}+b\log_{2}\frac{b}{d+b}
\label{cconst}
\end{eqnarray}
As already pointed out, the time-evolved state $\rho_{rs}(t)$ has the form given in equation (\ref{DEN2})
with the diagonal elements $a$, $b$ and $d$ being independent of time. From (\ref{cvalues}) and 
(\ref{redvalue}), $S\left(\rho_{B}\right)$ is thus independent of time. The other terms in equation 
(\ref{cconst}) are also independent of time since they involve only the elements $a$, $b$ and $d$.

\noindent \textit{Bit-phase-flip Channel.}         
Fig.\ref{bpf1} shows the plots of $I\left(\rho_{AB}\right)$ (solid line), $C\left(\rho_{AB}\right)$ 
(dashed line) and $Q\left(\rho_{AB}\right)$ (dotted line) as a function of $p$. Again, there is a sudden 
change, as in the case of the phase flip channel, in the decay dynamics of $C\left(\rho_{AB}\right)$
and $Q\left(\rho_{AB}\right)$ at $p=p_{sc}$ but in this case the two plots do not cross each other 
but touch at a single point $p=p_{sc}$. 

The dynamical features for the different quantum channels have been reported earlier \cite{maziero2}
for the class of states with $a=d$ in the reduced density matrix of equation (\ref{DEN2}), i.e., 
$c_{4}=0$ in equation (\ref{cvalues}). In our study, the reduced density matrix has the form shown in 
equation (\ref{DEN2}) with $a\neq d$, i.e., $c_{4}\neq 0$. The form corresponds to that of the two-qubit 
reduced density matrix obtained from the ground state of the TIM in 1d. In this case each quantum channel 
is distinguished by a specific type of dynamics. In Ref. \cite{maziero2}, the parameters 
$c_{1}$, $c_{2}$ and $c_{3}$ are free and different types of dynamics occur in different 
parameter regions corresponding to the same quantum channel.

\begin{figure}
\begin{center}
  \includegraphics[scale=0.6]{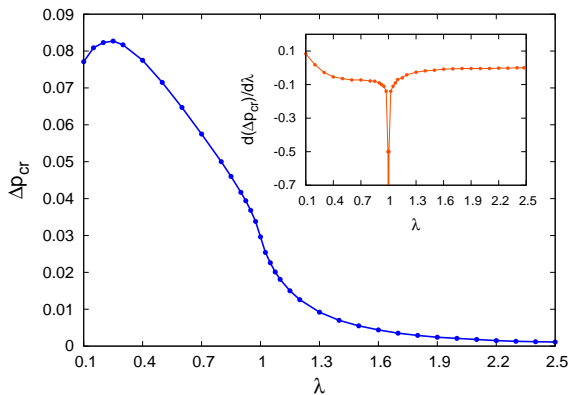}
\par\end{center}
\caption{(Color online) Variation of $\Delta p_{cr}=p_{cr_{2}}-p_{cr_{1}}$ with $\lambda$;
(inset) The first derivative of $\Delta p_{cr}$ w.r.t $\lambda$ diverges in the negative 
direction as the QCP $\lambda_{c}=1$ is approached}
\label{pf3}
\end{figure}

We now present some totally new results which have not been reported earlier. Fig.\ref{pf2} shows 
plots of $p_{cr_{1}}$ (solid line)  and $p_{cr_{2}}$ (dashed line) versus $\lambda$, the 
parameter appearing in the TIM Hamiltonian (equation (\ref{TIM})), in the case of the phase-flip channel. 
The inset of the Figure shows that the first derivative of $p_{cr_{1}}$ and $p_{cr_{2}}$ (both 
of which depend on $\lambda$)  w.r.t. the parameter $\lambda$ diverges as the QCP $\lambda_{c}=1$ is 
approached. The observation identifies a quantity  which provides the signature of a QPT occurring 
in a system subjected to decoherence under Markovian time evolution. Fig.\ref{bpf2} shows a variation 
of $p_{sc}$ with $\lambda$ for the phase-flip (dashed line) and the bit-phase-flip (solid line)
channels. Again, the inset shows that the first derivative of $p_{sc}$ w.r.t $\lambda$ diverges 
as the QCP $\lambda_{c}=1$ is approached. Fig.\ref{pf3} exhibits the plot of $\Delta p_{cr}
=p_{cr_{2}}-p_{cr_{1}}$ versus $\lambda$ in the case of the phase-flip channel. The inset shows that 
$\frac{d\left(\Delta p_{cr}\right)}{d\lambda}$ diverges in the negative direction as the QCP 
$\lambda_{c}=1$ is approached. We remind ourselves that when $p$ falls in the interval $\Delta p_{cr}$,
the quantum correlations $Q\left(\rho_{AB}\right)$ are larger in magnitude than the classical 
correlations $C\left(\rho_{AB}\right)$. Fig.\ref{pf3} is an outcome of the results of Fig.\ref{pf2}
as $\Delta p_{cr}=p_{cr_{2}}-p_{cr_{1}}$. In summary, the first derivative of any one of the quantities $p_{cr_{1}}$,
$p_{cr_{2}}$, $p_{sc}$ and $\Delta p_{cr}$ w.r.t $\lambda$ signals a quantum critical point transition.
The quantities correspond to states in which the quantum correlations are either equal to or 
greater than the classical correlations. At $p=p_{cr_{1}}$ and $p=p_{cr_{2}}$, 
$Q\left(\rho_{AB}\right)=C\left(\rho_{AB}\right)=I\left(\rho_{AB}\right)/2$ which is characteristic of
pure states with $Q\left(\rho_{AB}\right)$ being equal to the entropy of entanglement \cite{henderson,winter}.
The amplitude damping and bit-flip channels do not have these features. The appearance of singularities 
in the first derivatives of the quantities $p_{cr_{1}}$, $p_{cr_{2}}$, $p_{sc}$ and $\Delta p_{cr}$ 
with respect to the tuning parameter as the quantum critical point is approached can be explained 
by the fact that these quantities depend on two-spin correlation functions which exhibit 
a similar property close to the critical point. The non-trivial aspect arises from the identification 
of appropriate quantities associated with the dynamics of correlations which provide clear 
signatures of QPTs.

\section{Concluding remarks}
The TIM in 1d is a prototypical example of a quantum system exhibiting a QPT. The many body ground 
state has both classical and quantum correlations. The QD provides a quantitative measure of the quantum 
correlations in a two-qubit state. In this paper, we consider a two-qubit state described by the 
reduced density matrix obtained from the ground state of the TIM in 1d. The two-qubit state 
undergoes Markovian time evolution, described by the Kraus operator formalism, due to the local 
interactions of the qubits with independent environments. We consider the quantum channels, 
representing the interactions, to be of four types: amplitude damping, bit-flip, phase-flip and 
bit-phase-flip. The dynamics of the classical and quantum correlations exhibit distinctive 
features for each quantum channel. These features have been reported in an earlier study \cite{maziero2}
for a different class of initial states. In our study, we have not found evidence of another type
of dynamical behaviour mentioned in \cite{maziero2}, namely, that the classical correlations, 
$C\left(\rho_{AB}\right)$, are independent of time throughout the parametrized time interval 
$p$ whereas the QD, $Q\left(\rho_{AB}\right)$, decreases monotonically and becomes zero in the 
asymptotic limit $p\rightarrow 1$. The time evolution of the reduced density matrix of the TIM in 1d 
further does not exhibit the interesting dynamical behaviour described in \cite{mazzola}, namely, 
the existence of intervals of parametrized time when  $C\left(\rho_{AB}\right)$ and 
$Q\left(\rho_{AB}\right)$ are individually frozen. The most significant result of our study is the 
identification of quantities associated with the dynamics of the classical and quantum correlations 
which diverge as the QCP of the TIM in 1d, $\lambda_{c}=1$, is approached thus providing a 
distinctive signature of a QPT from a different perspective. The generalization of the result 
to other model systems exhibiting QPTs will certainly be of significant interest. In the present study, 
we have restricted our attention to Markovian time evolution. The more general case of 
non-Markovian time evolution can be investigated only after the appropriate calculational scheme 
for a system of interacting qubits is developed \cite{pal}.


\begin{thebibliography}{40}
 \bibitem{amico} L. Amico, R. Fazio, A. Osterloh, V. Vedral, Rev. Mod. Phys. \textbf{80}, 517 (2008)
 \bibitem{lewenstein} M. Lewenstein, A. Sanpera, V. Ahufinger, B. Damski, A. Sen, U. Sen, Adv. 
                      Phys. \textbf{56}, 243 (2007)
 \bibitem{horodecki} R. Horodecki, P. Horodecki, M. Horodecki, K. Horodecki, Rev. Mod. Phys. \textbf{81}, 
                     865 (2009)
 \bibitem{ollivier} H. Ollivier, W. H. Zurek, Phys. Rev. Lett. \textbf{88}, 017901 (2001)
 \bibitem{henderson} L. Henderson, V. Vedral, J. Phys. A: Math. Gen. \textbf{34}, 6899 (2001)
 \bibitem{zurek} W. H. Zurek, Rev. Mod. Phys. \textbf{75}, 715 (2003)
 \bibitem{datta} A. Datta, A. Shaji, C. M. Caves, Phys. Rev. Lett. \textbf{100}, 050502 (2008)
 \bibitem{lanyon} B. P. Lanyon, M. Bartieri, M. P. Almeida, A. G. White, Phys. Rev. Lett. \textbf{101}, 
                  200501 (2008)
 \bibitem{luo} S. Luo, Phys. Rev. A \textbf{77}, 042303 (2008)
 \bibitem{sarandy} M. S. Sarandy, Phys. Rev. A \textbf{80}, 022108 (2009)
 \bibitem{ali} M. Ali, A. R. P. Rau, G. Alber, Phys. Rev. A \textbf{81}, 042105 (2010)
 \bibitem{fanchini} F. F. Fanchini, T. Werlang, C. A. Brasil, L. G. E. Arruda, A. O. Caldeira, 
                    Phys. Rev. A \textbf{81}, 052107 (2010)
 \bibitem{maziero1} J. Maziero, T. Werlang, F. F. Fanchini, L. C. C\'{e}leri, R. M. Serra, Phys. 
                   Rev. A \textbf{81}, 022116 (2010)
 \bibitem{almeida} M. P. Almeida et al., Science \textbf{316}, 579 (2007)
 \bibitem{werlang1} T. Werlang, S. Souza, F. F. Fanchini, C. Villas Boas, Phys. Rev. A \textbf{80}, 024103
                   (2009)
 \bibitem{maziero2} J. Maziero, L. C. C\'{e}leri, R. M. Serra, V. Vedral, Phys. Rev. A \textbf{80}, 
                    044102 (2009)
 \bibitem{mazzola} L. Mazzola, J. Piilo, S. Maniscalco, Phys. Rev. Lett. \textbf{104}, 200401 (2010)
 \bibitem{pal} A. K. Pal, I. Bose, J. Phys. B: At. Mol. Opt. Phys. \textbf{44}, 045101 (2011)
 \bibitem{ferraro} A. Ferraro, L. Aolita, D. Cavalcanti, F. M. Cucchietti, A. Acin, Phys. Rev. 
                   A \textbf{81}, 052318 (2010)
 \bibitem{xu} J.-S. Xu, X.-Y. Xu, C.-F. Li, C.-J. Zhang, X.-B. Zou, G.-C. Guo, Nat. Commun. \textbf{1}, 
              7 (2010)
 \bibitem{auccaise} R. Auccaise et al., Phys. Rev. Lett. \textbf{107}, 140403 (2011)
 \bibitem{dutta} A. Dutta, U. Divakaran, D. Sen, B. K. Chakrabarti, T. F. Rosenbaum, 
                 G. Aeppli, arXiv:1012.0653v1 [cond-mat.stat-mech] (2011), 
                 to appear in Rev. Mod. Phys. 
 \bibitem{osborne} J. Osborne, M. A. Nielsen, Phys. Rev. A \textbf{66}, 032110 (2002)
 \bibitem{osterloh} A. Osterloh, L. Amico, G. Falci, R. Fazio, Nature \textbf{416}, 608 (2002)
 \bibitem{dillenschneider} R. Dillenschneider, Phys. Rev. B \textbf{78}, 224413 (2008)
 \bibitem{werlang2} T. Werlang, G. A. P. Ribeiro, G. Rigolin, Phys. Rev. A \textbf{83}, 062334 (2011)
 \bibitem{syljuasen} O. F. Sylju\aa{}sen, Phys. Rev. A \textbf{68}, 060301 (R) (2003) 
 \bibitem{barouch} E. Barouch, B. M. McCoy, Phys. Rev. A \textbf{3}, 786 (1971)
 \bibitem{pfeuty} P. Pfeuty, Ann. Phys. \textbf{57}, 79 (1970)
 \bibitem{nielsenchuang} M. A. Nielsen, I. L. Chuang in Quantum Computation and Quantum Information
                         (Cambridge University Press, Cambridge, England, 2000)
 \bibitem{salles}A. Salles et. al., Phys. Rev. A \textbf{78}, 022322 (2008)
 \bibitem{winter} B. Groisman, S. Popescu, A. Winter, Phys. Rev. A \textbf{72}, 032317 (2005) 
\end{thebibliography}
\end{document}